# Study of $CoFe_2O_4/CoFe_2$ nanostructured powder

E. S. Ferreira[1], E. F. Chagas[1], A. P. Albuquerque[1], R. J. Prado[1] and E. Baggio-Saitovitch[2]

[1]Instituto de Física, Universidade Federal de Mato Grosso, 78060-900 Cuiabá-MT, Brazil.

[2]Centro Brasileiro de Pesquisas Físicas, Rua Xavier Sigaud, 150 Urca Rio de Janeiro, Brazil

## Abstract

We report an experimental study of the $CoFe_2O_4/CoFe_2$ nanocomposite, a nanostructured material formed by hard ($CoFe_2O_4$) and soft ($CoFe_2$) magnetic materials. The precursor material, cobalt ferrite ($CoFe_2O_4$), was prepared using the conventional stoichiometric gel-combustion method. The nanocomposite material was obtained by reducing partially the precursor material using activated charcoal as reducing agent in air and argon atmospheres, at 800 and 900°C respectively. The magnetic hysteresis loops demonstrate that, in general, prepared nanocomposite samples display single magnetic behavior, indicating exchange coupling between the soft and hard magnetic phases. However, for nanocomposite samples prepared at higher temperatures, the hysteresis measurements show steps typical of two-phase magnetic behavior, suggesting the existence of two non-coupled magnetic phases. The studied nanocomposites presented coercivity ($H_C$) of about 0.7 kOe, which is considerably lower than the expected value for cobalt ferrite. A huge increase in $H_C$ (>440%) and maximum energy product (about 240%) was obtained for the nanocomposite after high energy milling processing.

## Introduction

The magnetic ferrites like $M^{2+}Fe_2^{3+}O_4$ (M = Ni; Co, Fe, Li, Mn, Zn, etc.) have been used for several applications such as high-density magnetic storage [1], electronic devices, biomedical applications [2-4], permanent magnets [5] and hydrogen production [6]. Among the hard ferrites the $CoFe_2O_4$ (cobalt ferrite) plays an important role, presenting promising characteristics such as high magnet-elastic effect [7], chemical stability, electrical insulation, moderate saturation magnetization ($M_S$), tunable coercivity ($H_C$) [8-11] and thermal chemical reduction [6, 12, 13]. However, for permanent magnet applications the parameters $M_S$ and $H_C$ are of fundamental importance, defining the quantity called as the figure of merit for permanent magnets, the energy product $(BH)_{max}$. This quantity tends to increase with increasing of both quantities. The $(BH)_{max}$ is an energy density (independent of the mass) that can be simplified as a measure of the maximum amount of magnetic energy stored in a magnet.

The tunable behavior of coercivity in cobalt ferrite allows the increase in $H_C$ [8, 11], and numerous methods as thermal annealing [9], capping [10] and mechanical milling treatment [8, 11] have been used to this purpose. For powdered cobalt ferrite materials, the highest coercivity achieved so far is 9.5 kOe, reported by Limaye *et al.* [10] through capping the nanoparticles with oleic acid. However, in that work, saturation magnetization decreased to 7.1 emu/g, a value about 10 times less than that expected for uncapped nanoparticles. On the other hand, Liu *et al.* [11] Ponce *et al.* [8] and Galizia *et al.* [14] obtained high coercivity $CoFe_2O_4$ (5.1, 4.2 and 3.7 kOe respectively) with relatively small decrease in $M_S$ using high energy mechanical milling.

Furthermore, the thermal chemical reduction enables us to increase the $M_S$ through the partial conversion of hard ferrimagnetic compound $CoFe_2O_4$ into the soft ferromagnetic intermetallic alloy iron cobalt ($CoFe_2$), producing the $CoFe_2O_4/CoFe_2$ exchange coupled nanocomposite [15-17]. The hysteresis curve for a nanocomposite with effective magnetic coupling should have a single material magnetic behavior (called by Zeng *et al.* [18] as single-phase-like behavior) with $M_S$ and $H_C$ between the values expected to $CoFe_2O_4$ ($M_S$ and $H_C$ about 70 emu/g and 1.0 kOe) and $CoFe_2$ ($M_S$ about 230 emu/g, though very small $H_C$). This interesting combination, of a hard magnetic material ($CoFe_2O_4$) with a soft one ($CoFe_2$) into an exchange coupled nanocomposite, presents enormous potential to hard magnetic applications.

In a previous work [13], of our group, hard/soft $CoFe_2O_4/CoFe_2$ nanocomposites were obtained by thermal treatment of a cobalt ferrite/carbon (activated charcoal) mixture in argon atmosphere, as indicated by equation below:

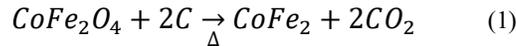

$$CoFe_2O_4 + 2C \xrightarrow{\Delta} CoFe_2 + 2CO_2 \qquad (1)$$

The symbol $\Delta$ indicates that thermal energy is necessary in the process.

According to the reduction process indicated in eq. (1), two moles of carbon are enough to convert all $CoFe_2O_4$ in $CoFe_2$ but, in practice, this does not happen. In this previous work [13], the reduction process was done using a powder of the mixture (cobalt ferrite plus carbon) in air and/or controlled inert atmosphere (argon). In all cases the mastering of the process was difficult and, in a sample treatment in argon atmosphere with 2 molar of C, only about 40% of $CoFe_2O_4$ was converted in $CoFe_2$. The main reason making difficult the control of the reduction process was attributed to the reaction of carbon with oxygen in air or with residual oxygen in the "inert" atmosphere [13].

Trying to solve these problems, in this work, we used a slightly different procedure to obtain the $CoFe_2O_4/CoFe_2$ nanocomposites. Here, the mixture of cobalt ferrite with activated charcoal was pressed to prepare a disk, aiming to avoid the contact between internal mixture (inside the disk) with the atmosphere of the furnace during the thermal treatment. This method could facilitate the control of $CoFe_2$ content in the nanocomposite $CoFe_2O_4/CoFe_2$. Moreover, the same milling procedure used previously to pure nanostructured cobalt ferrite [8] was performed in this work to increase the $H_C$ of the $CoFe_2O_4/CoFe_2$ nanocomposite, with excellent results.

## Experimental

The nanostructured cobalt ferrite precursor material was prepared using a conventional gel-combustion method as described in reference [19]. High-purity (99.9%) raw compounds were used. Cobalt nitrate and iron nitrate (VETEC, Brazil) were dissolved in 450 ml of distilled water in a ratio corresponding to the selected final composition. Glycine (VETEC, Brazil) was added in a proportion of one and half moles per mole of metal atoms, and the pH of the solution was adjusted with ammonium hydroxide (25%, Merck, Germany). The pH was tuned as closely as possible to 7, taking care to avoid precipitation. The resulting solution was concentrated by evaporation using a hot plate at 300 °C until a viscous gel was obtained. This hot gel finally burnt out as a result of a vigorous exothermic reaction. The system remained homogeneous throughout the entire process and no precipitation was observed. Finally, the as-reacted material was calcined in air at 700 °C for 2 h in order to remove the organic residues.

The nanocomposite was obtained mixing cobalt ferrite with activated charcoal for three different ratios 1:1; 1:2 and 1:5 (molar mass of $CoFe_2O_4$:C). The mixtures were pressed to form small disks of 10 mm diameter and about 1 mm thickness, thermally treated at 800 °C in air and at 900 °C in argon atmosphere. Aiming to obtain a totally reduced sample (transforming all $CoFe_2O_4$ into $CoFe_2$) a mixture powder with excessive quantity of carbon (1:24) was also prepared and thermally treated in a tubular furnace at 900 °C in argon atmosphere. For elucidation, the sample naming used in this work follow a simple labelling rule, for example, the name CFO-5C-800 indicate that 5 moles of charcoal was used during thermal reduction process of the sample and that it was thermally treated at 800 °C.

A Spex 8000 high-energy mechanical ball miller, with 6 mm diameter zirconia balls, was employed for the milling processing of all samples, aiming exclusively to increase their $H_C$. The processing time was 1.5 h for all samples, using ball/sample mass ratio of about 1/7. Detailed milling conditions are described in Ponce *et al.* [8].

The crystalline phases of the nanocomposite were identified by X-ray diffraction (XRD), using a Shimadzu XRD-6000 diffractometer installed at the *Laboratório Multiusuário de Técnicas Analíticas* (*LaMuTA*/ UFMT–Cuiabá-MT– Brazil). It is equipped with graphite monochromator and conventional Cu tube (0.154184 nm), and works at 1.2 kW (40 kV, 30 mA), using the Bragg-Brentano geometry. Magnetic measurements (hysteresis loops at 300 and 50 K) were carried out using a vibrating sample magnetometer (VSM) model VersaLab Quantum Design, installed at CBPF, Rio de Janeiro, Brazil.

## Results and discussion

It is well known that the magnetic behavior of $CoFe_2O_4$ is quite different from that found for $CoFe_2$. The former is a hard ferrimagnetic material with maximum $M_S$ of about 70 emu/g while the second is known to be a soft ferromagnet with $M_S$ of about 230 emu/g [20]. Consequently, in the case of nanocomposite formation presenting magnetic coupling, one can assume an intermediate magnetic behavior that depends of the relative amount of CoFe$_2$ formed during the reduction process and, also, of the microstructure of the material.

The hysteresis curves obtained for samples CFO-5C-800; CFO-2C-800; CFO-1C-800; CFO-5C-900; CFO-2C-900; CFO-1C-900 are shown in the figure 1. Measurements reveal clear differences between $M_S$ of the samples prepared at 800 °C, as seen in figure 1(A). However, coercivity values obtained are very close and the shape of the hysteresis loop is quite similar. Samples prepared at 900 °C also present similar $H_C$ but different $M_S$ values. On the other hand, the shape of the hysteresis loop obtained for sample CFO-5C-900 is different from those obtained for the two other samples prepared at 900 °C, as shown in figure 1(B).

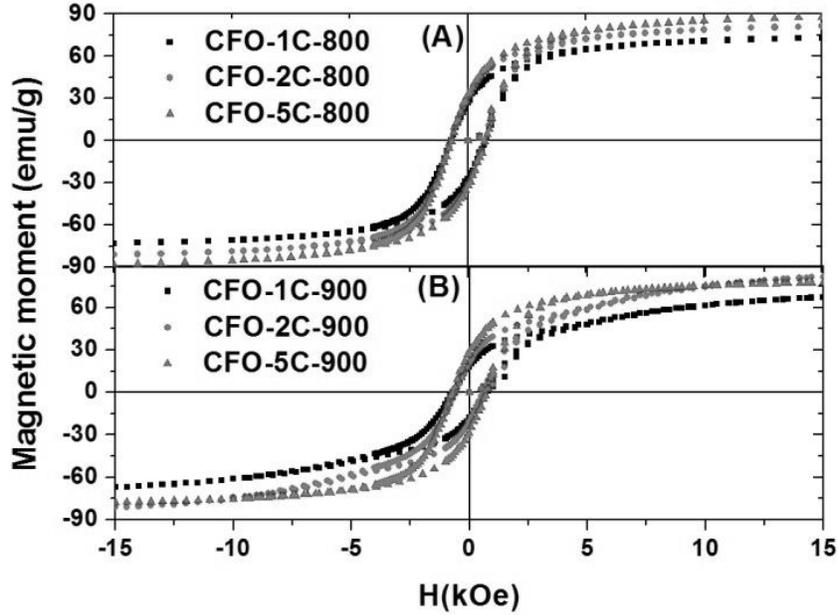

Figure 1 – Hysteresis curves at room temperature for samples treated at (A) 800 °C and (B) 900 °C.

There are only two reasonable possibilities to explain the different $M_S$ values obtained for the samples presented in figure 1: (i) an intense cationic redistribution caused by the thermal process used to prepare the sample or (ii) the formation of CoFe$_2$O$_4$/CoFe$_2$ composite.

To understand how the cationic redistribution could affect the $M_S$ is necessary to know the cobalt ferrite crystalline structure and its distribution of magnetic ions. The CoFe$_2$O$_4$ is a ferrimagnetic material that has an inverse spinel structure with three magnetic sites per formula unit. The CoFe$_2$O$_4$ structure can be summarized by [21]:

$$(Co^{2+}_{1-i}Fe^{3+}_i)[Co^{2+}_i Fe^{3+}_{2-i}]O^{2-}_4 \qquad (2)$$

In this representation, the round and the square brackets indicate A (tetrahedral coordination of four O$^{2-}$ anions site) and B sites (two octahedral coordination of six O$^{2-}$ anions sites), respectively, and $i$ (the degree of inversion) describes the fraction of the tetrahedral sites occupied by Fe$^{3+}$ cations. The ideal inverse spinel structure has $i = 1$ and the normal spinel has $i = 0$. A mixed spinel structure present $i$ values between 0 and 1.

In an ideal inverse spinel cobalt ferrite, half of the Fe$^{3+}$ cations (magnetic moment of 5μ$_B$) occupy the A-sites and the other half the B-sites, together with Co$^{2+}$ cations (see that there are two B sites per formula unit). Since the magnetic moments of the ions in the A and B sites are aligned in an anti-parallel way, there is no magnetic contribution of the $Fe^{3+}$ cations in this case. Therefore, the net magnetic moment of the ideal inverse spinel cobalt ferrite is due exclusively to the $Co^{2+}$ cations (magnetic moment of $3\mu_B$). However, normally, the cobalt ferrite presents a mixed spinel structure and, consequently, there are $Co^{2+}$ and $Fe^{3+}$ cations in both tetrahedral and octahedral sites, i.e. , when $i$ decreases toward 0 (normal spinel), it means swap of $Fe^{3+}$ cations from site A with $Co^{2+}$ cations from site B, increasing the net magnetic moment of the material and, consequently, promoting an increase of the $M_S$. Therefore, the magnetic behavior of the $CoFe_2O_4$, as saturation magnetization, is strongly affected by structural changes and/or chemical disorder/substitution.

As an example, to an ideal inverse spinel ($i = 1$) a magnetic moment of 3.0 $\mu_B$ per formula unit (equivalent to $M_S$ =71.4 emu/g) is expected. For sample CFO-5C-800 we obtained $M_S$ = 90 emu/g, that is equivalent to a magnetic moment of 3.8 $\mu_B$ per formula unit and $i = 0.8$, which means a swap of 20% of the Co cations in the material. This value to $i$ sounds reasonable, however highest values of $M_S$ found in the literature were 83.1 and 83.12 emu/g by Sato Turtelli et al. [21] and Kumar et al. [22], respectively. These literature values are equivalent to $\mu = 3.5$ $\mu_B$ per formula unit and $i = 0.88$ (12% of Co cations were swapped by Fe). For this reason we do not consider the cationic redistribution is responsible for the increase in $M_S$. Thus, after thermal chemical reduction, for the samples studied in this work, we consider the formation of $CoFe_2O_4/CoFe_2$ nanocomposite as responsible for the increase in $M_S$.

Using the extrapolation to zero of the M versus 1/H plot to estimate the $M_S$ values of each sample, and considering the $M_S$ expected to pure inverse spinel cobalt ferrite (71.4 emu/g) and pure $CoFe_2$ (230 emu/g) [20], one can estimate the content of $CoFe_2$ in the composite (see table 1). We do not consider the effect of canted magnetic moment from surface cations.

Table 1 – Magnetic parameters of the samples, obtained at room temperature.

| Sample | $M_S$ (emu/g) | $M_R$ (emu/g) | $M_R/M_S$ | $H_C$ (kOe) | Content of $CoFe_2$ (%) | $(BH)_{max}$ (MGOe) |
|---|---|---|---|---|---|---|
| CFO-5C-800 | 90 | 21 | 0.23 | 0.7 | 13 | 0.35 |
| CFO-2C-800 | 84 | 29 | 0.35 | 0.7 | 9 | 0.32 |
| CFO-1C-800 | 75 | 18 | 0.24 | 0.7 | 3 | 0.26 |
| CFO-5C-900 | 79 | 29 | 0.37 | 0.7 | 6 | 0.27 |
| CFO-2C-900 | 87 | 21 | 0.24 | 0.6 | 11 | 0.15 |
| CFO-1C-900 | 72 | 18 | 0.25 | 0.7 | 2 | 0.14 |

The $CoFe_2$ content in the nanocomposite was found to be small in all cases, indicating as false our initial assumption that carbon of the mixture inside the disk reacts only with oxygen from $CoFe_2O_4$ (generating $CoFe_2$). Two reasons can cause this effect: (i) residual oxygen inside the mixture sample disk reacted with the carbon and (ii) certain content of carbon do not reacts with cobalt ferrite during the thermal treatment. The second option probably occurs due to the morphology of the material, which permits the formation of a $CoFe_2$ shell around the $CoFe_2O_4$ core of the grain, shielding it and limiting the reduction power of the activated charcoal during thermal treatment. In any case, this effect makes difficult to control the quantity of $CoFe_2$ in the nanocomposite $CoFe_2O_4/CoFe_2$.

The small amount of $CoFe_2$ formed in samples treated in both air and argon atmospheres, as indicated in table 1, suggests that changes in the atmosphere of the furnace do not cause significant differences on the $CoFe_2$ content of the composite.

The coercivity observed for all samples was very similar, and close to 0.7 kOe, a value smaller than those obtained by Cabral *et al.* [12] and Leite *et al.* [13]. In other words, this value is lower than that expected for a pure cobalt ferrite and higher than that expected for pure $CoFe_2$ and, consequently, in agreement with the formation of the composite $CoFe_2O_4/CoFe_2$.

To perform a more detailed investigation about the differences between the shapes of the hysteresis curves obtained from samples prepared at 900 °C, the hysteresis loops were also measured at low temperature (50 K). These measurements are shown in figure 2. The low temperature hysteresis enhances the visualization of the behavior observed at room temperature (300 K). For example, the hysteresis curves obtained for samples CFO-2C-900 and CFO-1C-900 present a typical two-phase magnetic behavior, and the hysteresis curve obtained for sample CFO-5C-900 presents a single-phase magnetic behavior, suggesting the coupling between the different magnetic phases present into the sample. The derivative of the hysteresis curve (first quadrant decreasing field only) confirms our assumption (figure 2C), because the two-phase magnetic behavior is characteristic of two magnetic phases coexisting without exchange coupling between them. Similar behavior was observed by Sun *et al.* [23].

Nanocomposites with exchange coupling or exchange spring present single magnetic phase behavior [15-17, 24, 25], however, these effects depend on the crystallite size of the compounds. Zeng *et al.* [18] showed the single magnetic phase behavior can vanish, to the same nanocomposite, depending of the relative size of the compounds, resulting in two-phase behavior.

Interestingly, the two-phase magnetic behavior was not observed in samples prepared at 800 °C, even in the low temperature hysteresis curves, see figure 2B. This result reinforces our assumption that nanoparticles coalescence is the driving force towards the two-phase magnetic behavior. More visible effects of coalescence are expected to samples treated at higher temperatures, with consequent increase in crystallite size.

An additional aspect observed in 50 K hysteresis loop is the huge increase in coercivity. This effect was also observed by other authors [13, 26]. However, it is more significant in samples with single magnetic behavior presenting $H_C$ close to 5kOe (see figures 2A and 2B). In samples presenting two-phase behavior the coercivities were considerably smaller, *i.e.*, 2.4 kOe for CFO-2C-900 and $H_C$ = 1.8 kOe for sample CFO-1C-900 (figure 2A). A work studying this effect is under development.

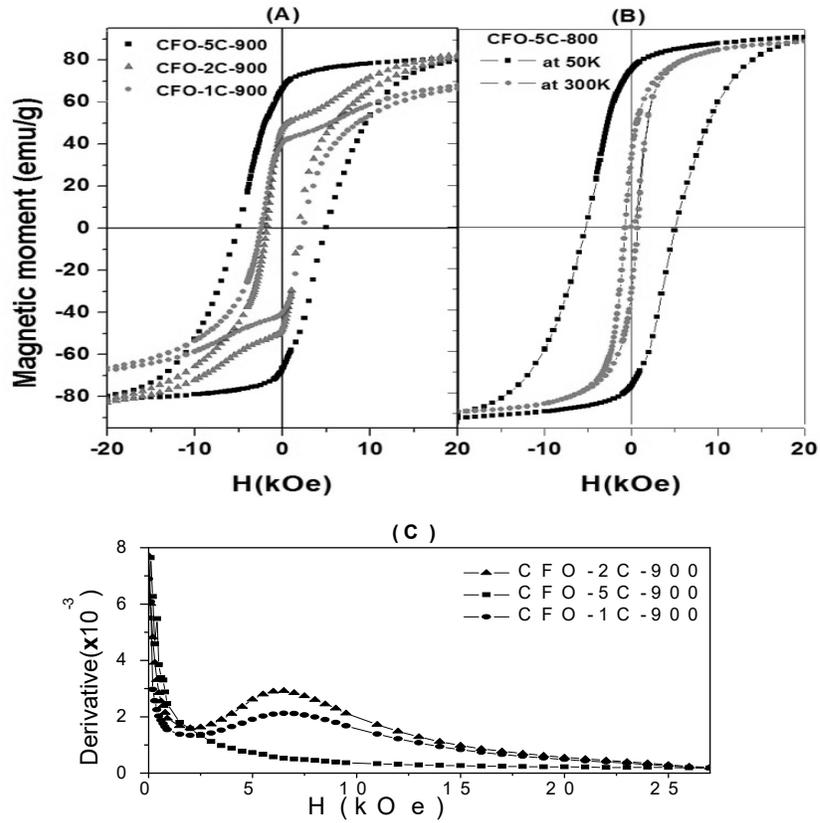

Figure 2 – Hysteresis curves (A) at 50K for samples treated at 900 °C and (B) at 300 and 50K to sample CFO-5C-800. (C) Derivative of the first quadrant (only decreasing field) M versus H, at 50K, for samples treated at 900 °C.

Liu and Ding [11] have shown that is possible to obtain a noteworthy increase in coercivity of cobalt ferrite powder via high-energy mechanical milling. Moreover, Ponce *et al.* [8] extended this method to nanostructured powders using specific milling parameters. Although, we could have had milled the cobalt ferrite to obtain a high coercivity $CoFe_2O_4$, but the thermal treatment at 800 or 900 °C used to obtain the nanocomposite $CoFe_2O_4/CoFe_2$ is known to decrease strain and structural defects and, consequently, also decreasing $H_C$ [19].

Aiming to obtain a similar effect for the nanocomposite $CoFe_2O_4/CoFe_2$, the samples CFO-2C-800 and CFO-1C-800 were milled using the same milling conditions described in the reference [8]. The result of the milling process to the sample CFO-2C-800 is shown in figure 4, a huge increase in $H_C$ of samples CFO-1C-800 (not shown) and CFO-2C-800 was obtained, showing that the effect obtained after milling in pure $CoFe_2O_4$ (a decrease of $M_S$ and an enormous increase of $H_C$) can also be achieved in the $CoFe_2O_4/CoFe_2$ nanocomposite. Specifically, in the case of sample CFO-2C-800, the coercivity at room temperature increased from 0.7 to 3.8 kOe and, as a consequence of the change in $H_C$, $(BH)_{max}$ increased from 0.32 to 1.1 MGOe at room temperature and to 2.6 MGOe at 50K. The magnetic parameters for the sample CFO-2C-800 before and after milling are presented in table 2. This is a very interesting result, since the high-energy milling procedure used in this work allows to obtain nanostructured powder with high $(BH)_{max}$.

Table 2 – Magnetic parameters obtained for sample CFO-2C-800, at 300 K before milling and at 300 and 50 K after milling process.

| Sample | T (K) | $M_S$ (emu/g) | $M_R$ (emu/g) | $M_R/M_S$ | $H_C$ (kOe) | $(BH)_{max}$ (MGOe) |
|---|---|---|---|---|---|---|
| CFO-2C-800 | 300 | 84 | 32 | 0.38 | 0.7 | 0.32 |
| CFO-2C-800(milled) | 300 | 60 | 34 | 0.57 | 3.8 | 1.1 |
| CFO-2C-800(milled) | 50 | 64 | 48 | 0.75 | 11.1 | 2.6 |

Figure 3 presents a clear decrease of $M_S$ after milling, this effect is associated to the decrease of the mean crystallite size due to the milling process [8, 11, 14]. The reduction of the crystallite size increases the number of surface magnetic ions and the canting effect of these surface ions is responsible for the decrease in $M_S$.

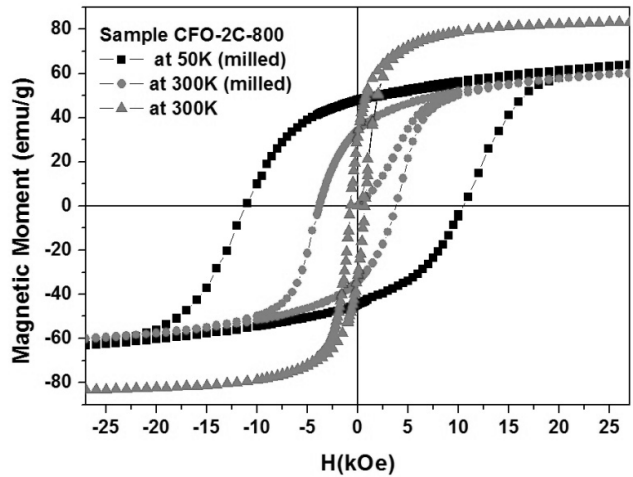

Figure 3 – Hysteresis curves for sample CFO-2C-800, obtained at 300 K before and after milling and at 50 K after milling process.

An additional interesting property was observed in the hysteresis loop at 50K (see figure 3). Milled $CoFe_2O_4/CoFe_2$ nanocomposites present $H_C$ more than twice higher than those of non-milled samples (coercivity changes from 5.0 to 11 kOe after milling). This enormous increase in $H_C$ with the decrease of temperature requires further investigation.

XRD experiments were also performed, in order to obtain diffraction patterns from cobalt ferrite samples before (pristine sample) and after thermal treatment at 800 and 900 °C in air and argon atmosphere respectively, as shown in figure 4. The inset of figure 4 presents the x-ray diffraction pattern of the pristine sample and a refinement using *Match* software (solid line). The only objective of this refinement is to show that pristine cobalt ferrite is monophasic.

The comparison between XRD patterns before and after thermal treatment at 800 °C show no visible significant differences, except for the narrowing of the peaks after thermal treatment. Such narrowing is usually related to the reduction in quantity of defects, improvements in the crystal lattice distortion around defects (local strain) and/or, consequently, to the increase of the mean crystallite size of the material [27].

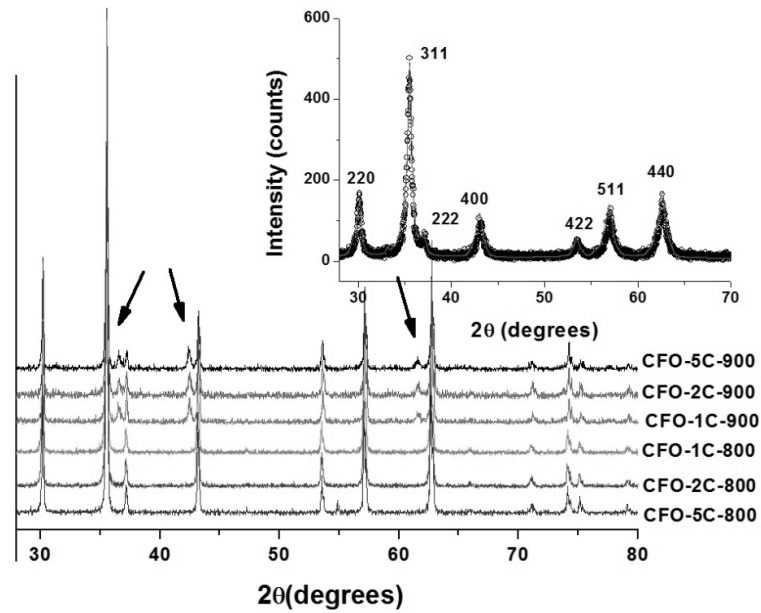

Fig. 4 - XRD patterns of several analyzed samples, as labeled.

The mean crystallite size of the cobalt ferrite phase (see table 3) was estimated by the Scherrer method for the (311) diffraction peak, at 35.5 ° in 2θ. The results, as expected, indicate that crystallite size increases with the temperature of the thermal treatment.

Table 3 – Mean crystallite size estimated from the Scherrer equation for the cobalt ferrite (311) XRD peak of the samples.

| Sample | D (nm) |
|---|---|
| Pristine | 12 |
| CFO-5C-800 | 39 |
| CFO-2C-800 | 34 |
| CFO-1C-800 | 38 |
| CFO-5C-900 | 57 |
| CFO-2C-900 | 45 |
| CFO-1C-900 | 49 |

However, samples thermally treated at 900 °C presented three new small diffraction peaks (indicated by arrows in figure 4). These peaks do not match with the XRD patterns expected for $CoFe_2$ [13], being identified as CoO crystalline phase (ICSD Card # 9865). The existence of these small peaks indicates phase separation due to a small lack in stoichiometry of the synthesized material. This small difference in stoichiometry is common in a chemical process of synthesis, due to the hygroscopic nature of the reagents used in the process and, if necessary, can be eliminated changing in a few percent the cobalt and/or iron nitrate masses used in the synthesis. Consequently, these small peaks do not indicate partial reduction of the material (formation of $CoFe_2$) and, also, the CoO phase does not contribute to hysteresis curve because it is antiferromagnetic [28].

Therefore, we associate the two-phase behavior to the increase in the mean crystallite size. This assumption agrees with the behavior of the mean crystallite size in function of the thermal treatment temperature obtained from Scherrer equation (table 3).

The absence of x-ray diffraction peaks related with $CoFe_2$ phase in the patterns shown in figure 1 agrees with results obtained by Zhang *et al.* [15]. These authors, using a similar process, could not observe the presence of $CoFe_2$ in the nanocomposite $CoFe_2/CoFe_2O_4$, by XRD, until the content reached about 30 %, relating the absence of $CoFe_2$ phase diffraction peaks to its poor crystallinity. Therefore, the absence of $CoFe_2$ diffraction peaks in the diffractograms shown in figure 4 does not exclude the formation of this phase.

For information, the thickness of the $CoFe_2$ shell around the nanostructured $CoFe_2O_4$ core was estimated for sample CFO-5C-800, that with the higher $CoFe_2$ content, supposing spherical $CoFe_2O_4$ nanoparticles with diameter equal to the mean crystallite size obtained from XRD measurements (see table 3) covered with a uniform $CoFe_2$ shell. Mass density of both $CoFe_2O_4$ and $CoFe_2$ crystalline phases were used. The result indicates that only a very thin $CoFe_2$ shell, having around 1nm, was formed. One has, also, to consider that (i) if disordered $CoFe_2$ shell and $CoFe_2O_4$ core can result in respectively larger and/or smaller values for the $CoFe_2$ shell thickness, reducing the importance of this consideration in a more precise calculation, (ii) a nanostructured material (as that studied in this work) can present much more relative surface/volume area than that of a perfect sphere, further reducing the estimated thickness of the $CoFe_2$ shell. Therefore, the thickness of the $CoFe_2$ shell must be less than the estimated 1 nm. This estimative is fully consistent with the absence of diffraction peaks from a highly disordered $CoFe_2$ phase.

Aiming to maximize the reduction of a $CoFe_2O_4$ during thermal treatment (transforming more $CoFe_2O_4$ into $CoFe_2$), a mixture powder with an extreme quantity of carbon (1:24) was also prepared and thermally treated in a tubular furnace at 900 °C in argon atmosphere. The XRD pattern of sample CFO-24-900 is shown in figure 5 together with the theoretical diffractogram expected for a pure $CoFe_2$ sample. The result suggests that something between an almost total and a total conversion of $CoFe_2O_4$ into $CoFe_2$ took place.

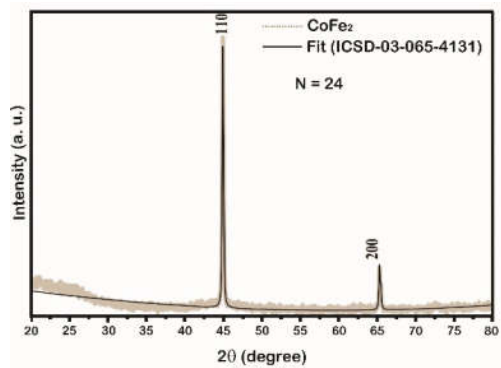

Figure 5 - XRD pattern obtained from sample CFO-24-900 and theoretical fit supposing $CoFe_2$ crystalline structure.

Finally, on the difficulty reported here, and in previous work [13], in attempting to master the reduction process, this last result indicate that a greater amount of carbon has to be used to achieve larger conversion efficiency, even in inert atmosphere. Future investigations are needed to improve reduction process efficiency of $CoFe_2O_4$ in $CoFe_2$, in order to evaluate the magnetic behavior of the nanocomposite with higher $CoFe_2$ content.

## Conclusion

We use a new procedure to obtain the $CoFe_2O_4/CoFe_2$ nanocomposite. Despite the XRD patterns suggesting that $CoFe_2$ is not formed, magnetic measurements indicate that nanocomposite are obtained, however with small amount. The preparation method of the nanocomposite at 900°C produces samples with two magnetic phase behavior and a third phase was observed the antiferromagnetic cobalt oxide CoO. This behavior was not observed in samples prepared at 800°C, indicating an exchange coupling between the magnetic phases.

The most important result of this work is to show that $H_C$ of the $CoFe_2O_4/CoFe_2$ nanocomposite can be increased using the same milling method employed previously to cobalt ferrite. As a consequence, we obtained a sample that reached an increase in $(BH)_{max}$ of about 240%, but future investigations are needed to evaluate the magnetic behavior of the nanocomposite with higher $CoFe_2$ content.

## Acknowledgments

The authors would like to thank the CAPES Brazilian funding agency to the master students grant. In addition, E. Baggio-Saitovitch acknowledges support from FAPERJ through several grants including Emeritus Professor fellow and CNPq for BPA and corresponding grants.

## References

1. V. Pillai, D. O. Shah, Synthesis of High-Coercivity Cobalt Ferrite Particles Using Water-in-Oil Microemulsions J. Magn. Magn. Mater. 163 (1996) 243-248.

2. M. Colombo, S. Carregal-Romero, M. F. Casula, L. Gutierrez, M. P. Morales, I. B. Bohm, J. T. Heverhagen, D. Prosperi, J. W. Parak, Biological applications of magnetic nanoparticles, Chem. Soc. Rev. 41 (2012) 4306-4334.

3. T.E. Torres, A.G. Roca, M.P. Morales, A. Ibarra, C. Marquina, M.R. Ibarra, G.F. Goya, Magnetic properties and energy absorption of $CoFe_2O_4$ nanoparticles for magnetic hyperthermia, J. Phys.: Conf. Ser. 200 (2010) 072101.

4. J.H. Lee, J.T. Jang, J.S. Choi, S.H. Moon, S.H. Noh, J.W. Kim, J.G. Kim, I.S. Kim, K. I. Park, J. Cheon, Exchange-coupled magnetic nanoparticles for efficient heat induction, Nat. Nanotechnol. 6 (2011) 418-422.

5. Y.C. Wang, J. Ding, J.B. Yi, B. H. Liu, T. Yu, Z.X. Shen, High-coercivity Co-ferrite thin films on (100) SiO2(100)-SiO2 substrate, Appl. Phys. Lett. 84 (2004) 2596-2598.

6. J.R. Scheffe, M.D. Allendorf, E.N. Coker, B.W. Jacobs, A.H. McDaniel, A.W. Weimer, Hydrogen Production via Chemical Looping Redox Cycles Using Atomic Layer Deposition-Synthesized Iron Oxide and Cobalt Ferrites, Chem. Mater. 23 (2011) 2030-2038.

7. C. Nlebedim, J.E. Snyder, A.J. Moses, D.C. Jiles, Dependence of the magnetic and magnetoelastic properties of cobalt ferrite on processing parameters, J. Magn. Magn. Mater. 322 (2010) 3938-3941.

8. A.S. Ponce, E.F. Chagas, R.J. Prado, C.H.M. Fernades, A.J. Terezo, E. Baggio-Saitovitch, High coercivity induced by mechanical milling in cobalt ferrite cobalt powders, J. Magn. Magn. Mater. 344 (2013) 182-187.

9. W.S. Chiua, S. Radiman, R. Abd-Shukor, M.H. Abdullah, P.S. Khiew, Tunable coercivity of CoFe2O4 nanoparticles via thermal annealing treatment, J Alloys Compd 459 (2008) 291–297.

10. M.V. Limaye, S.B. Singh, S.K. Date, D. Kothari, V.R. Reddy, A. Gupta, V. Sathe, R. J. Choudhary, S.K. Kulkarni, High coercivity of oleic acid capped CoFe2O4 nanoparticles at room temperature, J. Phys. Chem. B 113 (2009) 9070–9076.

11. B. H. Liu and J. Ding, Strain-induced high coercivity in CoFe2O4 powders, Appl. Phys. Lett. 88 (2006) 042506.

12. F.A.O. Cabral, F.L.A. Machado, J.H. Araujo, J.M. Soares, A.R. Rodrigues, A. Araujo, Preparation and magnetic study of the CoFe2O4-CoFe2 nanocomposite powders, IEEE Trans. Magn. 44 (2008) 4235-4238.

13. G.C.P. Leite, E.F. Chagas, R. Pereira, R.J. Prado, A.J. Terezo, E. Baggio-Saitovitch, Exchange coupling behavior in bimagnetic CoFe2O4/CoFe2 J. Magn. Magn. Mater. 324 (2012) 2711-2016.

14. P. Galizia, M. Cernea, V. Mihalache, L. Diamandescu, G. Maizza, C. Galassi, Easy batch-scale production of cobalt ferrite nanopowders by two-step milling: Structural and magnetic characterization, Mater. Design 130 (2017) 327-335.

15. Y. Zhang, B. Yan, Jun O. Yang, B. Zhu, S. Chen, X. Yang, Y. Liu, R. Xiong, Magnetic properties of core/shell-structured $CoFe_2/CoFe_2O_4$ composite nano-powders synthesized via oxidation reaction, Ceram. Int. 41 (2015) 11836–11843.

16. J.M. Soares, F.A.O. Cabral, J.H. de Araújo, F.L.A. Machado, Exchange-spring behavior in nanopowders of $CoFe_2O_4$–$CoFe_2$, Appl. Phys. Lett. 98 (2011) 072502.

17. J.M. Soares, V.B. Galdino, O.L.A. Conceição, M. A. Morales, J.H. de Araújo, F.L.A. Machado, Critical dimension for magnetic exchange-spring coupled core/shell $CoFe_2O_4/CoFe_2$ nanoparticles J. Magn. Magn. Mater. 326 (2013) 81–84.

18. H. Zeng, J. Liu, J.P. Liu, Z.L. Wang, S. Sun, Exchange-coupled nanocomposite magnets by nanoparticle self-assembly, Nature 420 (2002) 395-398.

19. E. F. Chagas A.S. Ponce, R.J. Prado, G.M. Silva, J. Bettini, E. Baggio-Saitovitch, Thermal effect on magnetic parameters of high-coercivity cobalt ferrite, J. Appl. Phys. 116 (2014) 033901.

20. T.Ungar in "Industrial Applications of X-ray Diffraction"; eds: Chung & Smith. Marcel Dekker, New York, 2000. P. 847.

21. M. Mohan, V. Chandra, S.S. Manoharan, A new nano $CoFe_2$ alloy precursor for cobalt ferrite production via sonoreduction process, Curr. Sci. 94 (2008) 473–476.

22. R.Sato Turtelli, M.Atif, N.Mehmood, F.Kubel, K.Biernacka, W.Linert, R.Grössinger, Cz.Kapusta, M.Sikora, Interplay between the cation distribution and production methods in cobalt ferrite Mater. Chem. Phys. 132 (2012) 832-838.

23. K. Kumar, A.Sharma, Md. A. Ahmed, S. S. Mali, C. K. Hong, P. M. Shirage Morphology-controlled synthesis and enhanced energy product $(BH)_{max}$ of $CoFe_2O_4$ nanoparticles, New J. Chem. 42 (2018) 15793-15802.

24. X. Sun, Y.Q. Ma, Y.F. Xu, S.T. Xu, B.Q. Geng, Z.X. Dai, G.H. Zheng, Improved magnetic performance at low and high temperatures in non-exchange-coupling $CoFe_2O_4/CoFe_2$ nanocomposites, J Alloys Compd 645 (2015) 51-56.

25. N. J. O. Silva, A. Millan, F. Palacio, M Martins, T. Trindade, I Puente-Orench, J. Campos, Remanent magnetization in CoO antiferromagnetic nanoparticles, Phys. Rev. B 82 (2010) 094433–094440.

26. E.F. Kneller, R. Hawig, The exchange-spring magnet: a new material principle for permanent magnets, IEEE Trans. Magn. 27 (1991) 3588–3600.

27. R. Skomski, J.M.D. Coey, Giant energy product in nanostructured two phase magnets, Phys. Rev. B 48 (1993) 15812–15816.

28. K.Gandha, K. Elkins, N. Poudyal, J. P. Liu, Synthesis and characterization of $CoFe_2O_4$ nanoparticles with high coercivity, J. App. Phys. 117 (2015) 7A7361.